# Representing Brain Anatomical Regularity and Variability by Few-Shot Embedding

Lu Zhang[1*], Xiaowei Yu[1], Yanjun Lyu[1], Zhengwang Wu[2], Haixing Dai[3], Lin Zhao[3], Li Wang[2], Gang Li[2], Tianming Liu[3], Dajiang Zhu[1]

[1] Department of Computer Science and Engineering, University of Texas at Arlington
[2] Department of Radiology and BRIC, University of North Carolina at Chapel Hill
[3] Department of Computer Science and Bioimaging Research Center, University of Georgia
`lu.zhang2@mavs.uta.edu`

**Abstract.** Effective representation of brain anatomical architecture is fundamental in understanding brain regularity and variability. Despite numerous efforts, it is still difficult to infer reliable anatomical correspondence at finer scale, given the tremendous individual variability in cortical folding patterns. It is even more challenging to disentangle common and individual patterns when comparing brains at different neurodevelopmental stages. In this work, we developed a novel learning-based few-shot embedding framework to encode the cortical folding patterns into a latent space represented by a group of anatomically meaningful embedding vectors. Specifically, we adopted 3-hinge (3HG) network as the substrate and designed an autoencoder-based embedding framework to learn a common embedding vector for each 3HG's multi-hop feature: each 3HG can be represented as a combination of these feature embeddings via a set of individual specific coefficients to characterize individualized anatomical information. That is, the regularity of folding patterns is encoded into the embeddings, while the individual variations are preserved by the multi-hop combination coefficients. To effectively learn the embeddings for the population with very limited samples, few-shot learning was adopted. We applied our method on adult HCP and pediatric datasets with 1,000+ brains (from 34 gestational weeks to young adult). Our experimental results show that: 1) the learned embedding vectors can quantitatively encode the commonality and individuality of cortical folding patterns; 2) with the embeddings we can robustly infer the complicated many-to-many anatomical correspondences among different brains and 3) our model can be successfully transferred to new populations with very limited training samples.

## 1    Introduction

Accumulating evidence suggests that brain anatomical structure, e.g., cortical folding patterns, can reveal the underlying mechanisms of brain organizational architecture [1-3]. Thus, how to effectively characterize and represent cortical folding is of fundamen-tal importance in understanding brain complexity, including both regularity and individual variability. Prior research has focused primarily on two streams of anatomical representation. One is using local shape-based descriptors, such as curvature and its derivations. Despite being widely used, they can only depict small neighborhood of a cortical foci that may be difficult, or even impossible, to find a very accurate correspondence on other brains. The second way is to use global descriptors, such as gyrification index [4]. They can quantify the overall cortical convolution level of either entire cortical surface or preselected regions of interest (ROIs). Yet, these descriptors usually need to be averaged before conducting group-wise analysis in order to achieve statistical power. In addition, current methods (both local and global descriptors) do not provide a quantitative encoding of the regularity and individuality of brain anatomical structure, making it difficult to infer accurate correspondence upon extremely complicate cortical landscape [5]. Recent emergence of learning-based embedding methods offers great promise for transcending these limitations [6-10]. The core idea of embedding is to transfer the original data (e.g., brain anatomical structure) to a latent space that is represented by a group of learned embedding vectors. With these embedding vectors, we can quantitatively and flexibly characterize and represent group-wise consistency (commonality) and individual specific patterns (individuality). However, learning-based embedding tends to need more samples in the training step. For lifespan brain imaging dataset, for example, the populations at many early neurodevelopmental stages often have very limited sample size, comparing to adult data that can be more than 1,000 (e.g., HCP dataset). Hence, a new embedding strategy that can leverage the embeddings learned from the population with larger sample size and effectively transfer this knowledge together with the encoded regularity/variability to other populations with relatively smaller samples size is highly desired.

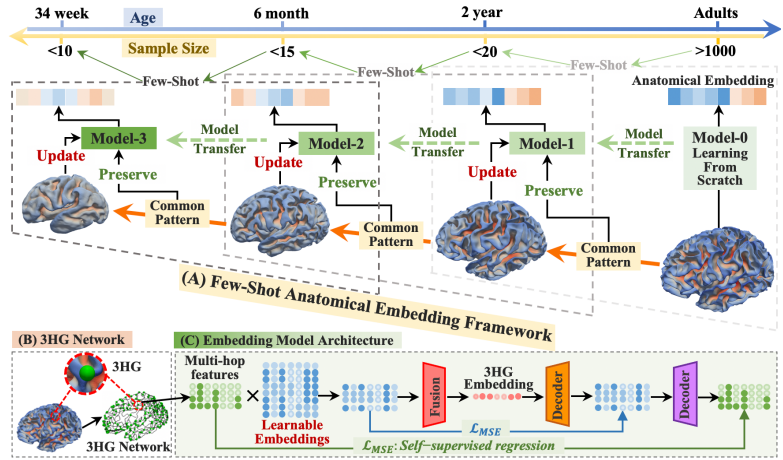

**Fig. 1.** (A) The overall architecture of the proposed framework. The embedding model (C) is trained from scratch by the adult group which has the most data samples and then transferred and adapted to other age groups with very limited data samples. (B) We used the 3HG network to describe the brain anatomy, where each 3HGs is the conjunction of gyri from three directions.

To achieve this goal, in this work, we proposed a learning-based few-shot anatomical embedding framework (Fig. 1 (A)) to encode the cortical folding patterns into a group of anatomically meaningful embeddings for different age groups. As the formation of the cortical folding is a continues process, the anatomical patterns of two close age groups share more commonalities. To take advantage of this kind of commonalities, we trained the model from scratch by the age group with the largest and sufficient sample size and then transferred and adapted the well-trained model to the next closer age group. The commonalities will be preserved, and the new patterns can be updated. We adopted 3-hinge (3HG) network (Fig. 1 (B)) [11-13] as substrate and designed an auto-encoder-based embedding framework to learn a common embedding vector for each 3HG's multi-hop feature: each 3HG can be represented as a combination of these feature embeddings via a set of individual specific coefficients to characterize individualized anatomical information. That is, the regularity of folding patterns is encoded into the embeddings, while the individual variations are preserved by the multi-hop combination coefficients. We applied our method on adult HCP and pediatric datasets with 1,000+ brains (from 34 gestational weeks to young adult). Our experimental results show that: 1) the learned embeddings can quantitatively encode the commonality and individuality of cortical folding patterns; 2) with the embeddings we are able to robustly infer the complicated many-to-many anatomical correspondences among different brains and 3) our model can be successfully transferred to new populations (e.g., early neurodevelopment) with very limited training samples (*code and data will be released*).

## 2    Methods

### 2.1    3HG Multi-Hop Feature Generation

For each cerebral hemisphere, the 3HGs were identified automatically via a pipeline [11] (details can be found in supplementary materials) and connected into a network by gyral hinges. Let $\mathcal{G} = (\mathcal{V}, \mathcal{E})$ denote the 3HG network, where $\mathcal{V} = \{v_1, v_2, \cdots, v_N\}$ is the set of $N$ 3HGs and $\mathcal{E} \subseteq \{\{v_i, v_j\} | v_i, v_j \in \mathcal{V}\}$ is the set of unweighted edges. Its adjacency matrix is denoted by $\mathcal{A} = [a_{i,j}] \in R^{N \times N}$, where $a_{i,j} = 1$ if there is a connection between $v_i$ and $v_j$, and $a_{i,j} = 0$ otherwise. We conducted ROI labeling via Destrieux Atlas [14] and divided the whole surface into 75 ROIs. Each 3HG was assigned an ROI label as node feature. We numerically represented the ROI labels by one-hot encoding, i.e., the $k^{th}$ label was denoted by $e_k \in R^{75}$ with 1 in the $k^{th}$ location and 0 elsewhere. Accordingly, 3HG- $i$ in ROI- $k$ can be denoted by $x_i = e_k$.

By far, the network of 3HGs can be represented by the adjacency matrix $\mathcal{A}$ and the feature matrix $\mathcal{X} = \{x_1; x_2; \cdots; x_N\} \in R^{N \times 75}$. Based on the two matrices, we defined the $l^{th}$ hop feature matrix of $\mathcal{G}$ as $\mathcal{A}^l \mathcal{X}$, where $\mathcal{A}^l$ is the $l^{th}$ power of $\mathcal{A}$. As the adjacency matrix $\mathcal{A}$ defines the direct connections between nodes, during the process of recurrently multiplying by itself, which is like the graph convolution operation, the undirected connections of further neighbors are propagated and gathered along with the direct connections. When multiplying $\mathcal{A}$ $l$ times, the features of the neighbors that can reach the center node by $l$ steps are congregated. As each row of $\mathcal{A}^l \mathcal{X}$ corresponds to one 3HG, the $l^{th}$ hop feature of 3HG-$i$ thereby can be denoted by $\boldsymbol{a}_i^l \mathcal{X}$, where $\boldsymbol{a}_i$ is the $i^{th}$ row of $\mathcal{A}$. Based on the above discussion, we further defined the multi-hop features of 3HG-$i$ as follows (taking $l$-hop features as an example):

$$F_{MH}^i = \left[ x_i; \ \boldsymbol{a}_i^1 \mathcal{X}; \ \boldsymbol{a}_i^2 \mathcal{X}; \ \cdots; \boldsymbol{a}_i^l \mathcal{X} \right] \in R^{(l+1) \times 75} \ , \tag{1}$$

where $\boldsymbol{a}_i^l \mathcal{X} = \sum_{j=1}^N a_{ij}^l x_j$, and $x_j \in \{e_k | k = 1, 2, \cdots, 75\}$, hence $\boldsymbol{a}_i^l \mathcal{X}$ can be represented by the group of one-hot label embeddings: $\boldsymbol{a}_i^l \mathcal{X} = \sum_{k=1}^{75} c_{ik}^l e_k$ with the multi-hop coefficient $c_{ik}^l$ to indicate the number of different $l$-step paths that available from 3HG-$i$ to the ROI-$k$. Organizing the multi-hop features in this manner, all the 3HGs are mapped into the same feature space defined by the group of ROI labels. By possessing a set of specific multi-hop coefficients $\{c_{ik}^l\}_{k=1}^{75}$, the variability of 3HGs can be represented. However, current discrete one-hot encoding of the ROI labels can be limited in robustness when representing the commonality and individual variability. Hence, we designed a learning based embedding method to learn anatomically meaningful multi-hop feature embeddings in section 2.2.

## 2.2 Learning Based Multi-Hop Feature Embedding

Our learning-based embedding framework (Fig. 1(C)) is designed in a self-supervised manner, with two-level encoding to hierarchically map the input multi-hop features to a latent representation as well as a two-level decoding that hierarchically reconstruct the original input from the latent representation. The encoding and decoding processes can be formulated as (2) and (3), respectively:

$$E_{MH}^i = \sigma\left( F_{MH}^i \cdot W^{Embedding} \right) \ \text{(2-a)} \qquad E_F^i = \sigma\left( (E_{MH}^i)^T \cdot W^{Fusion} \right) \ \text{(2-b)} \tag{2}$$

$$\hat{E}_{MH}^i = \left( \sigma\left( E_F^i \cdot W^{D_1} \right) \right)^T \ \text{(3-a)} \qquad \hat{F}_{MH}^i = \hat{E}_{MH}^i \cdot W^{D_2} \qquad \text{(3-b)} \tag{3}$$

where $T$ is the transpose operation, and $W^{Embedding} = \{w_1; w_2; \cdots; w_{75}\} \in R^{75 \times d}$ is the learnable feature embedding. In our setting, there are 75 ROIs and we initialized a learnable embedding vector $w_i \in R^d$ for each ROI. The input multi-hop features are firstly embedded via $W^{Embedding}$ hop by hop and the multi-hop embeddings $E_{MH}^i \in R^{(l+1) \times d}$ is generated. To further fuse the multi-hop embeddings into a single embedding vector that contains the completed multi-hop information, we conducted the second encoding by learnable combination parameters $W^{Fusion} \in R^{(l+1) \times 1}$ to fuse the multi-hop embeddings into the combined embedding vector $E_F^i \in R^{d \times 1}$.

Similar with the classical autoencoder, we used a symmetric design for the two-level decoding with the parameters $W^{D_1} \in R^{1 \times (l+1)}$ and $W^{D_2} \in R^{d \times 75}$, respectively. The first decoding reconstructs the hierarchical multi-hop embeddings from the combined embedding vector, which can help to ensure that the combined embedding vector $E_F^i$ has captured the completed information to restore the embeddings for each hop. Then, upon the restored multi-hop embeddings $\tilde{E}_{MP}^i$, the second decoding was applied to recover the multi-hop input features $\tilde{F}_{MH}^i$. We adopted the MSE loss to evaluate the two-level decoding and the objective function can be defined by (3):

$$\mathcal{L} = \alpha \left\| E_{MH}^i - \widehat{E}_{MH}^i \right\|_F^2 + \beta \left\| F_{MH}^i - \widehat{F}_{MH}^i \right\|_F^2 \qquad (4)$$

where $\alpha$ and $\beta$ are the regularization parameters to control the contribution of the two-level decoding. The whole model was trained in a self-supervised manner, which plays a key role in the learning process as it establishes the connection between the learnable embeddings with the input data without any bias introduced by supervised regular terms.

### 2.3 Few-Shot Strategy

In this work, the dataset includes four groups: adult, 2-year (2Y), 6-month (6M) and 34-week (34W). Based on Section 2.1, we generated multi-hop features of 3HGs for each age group. The anatomical features of the four groups are numerically represented in the same feature space. Through this way, the commonality shared by different age groups and group specific patterns can be encoded into the feature representations. We trained the whole model from scratch using adult group, which has the largest sample size, to endow the model good generalizability. When transferring the well-trained model to other three age groups, as the common patterns shared the same feature representation, it will generate relatively small back-propagated gradients and thus will not change the model much. The limited data samples will be fully utilized to learn the group specific patterns. To maximize the effectiveness when learning the commonality of folding patterns, models were transferred sequentially from one age group to the next closest group, since they tend to share the greatest consistency.

## 3 Results

### 3.1 Experimental Setting

**Data Setting.** In this work, we used structural MRI of 1,064 adults from Human Connectome Project (HCP) S1200 release. The detailed imaging parameters can be referred to [15]. We applied the same standard pre-processing procedures as in [16] for imaging data. In brief, pre-processing steps include brain skull removal, tissue segmentation and cortical surface reconstruction by FreeSurfer package [17]. To demonstrate the effectiveness of our few-shot strategy, we only used pediatric structural MRI of 12/10/8 subjects in 2-year/6-month/34-week groups from NDA and dHCP datasets. All pediatric images were processed with an infant-dedicated pipeline (http://www.ibeat.cloud/). Destrieux parcellation [14] was used to conduct ROI labeling for all of the four age groups. After preprocessing, 500/64/500 training/validation/testing splitting was adopted for the adult group. And for the other three groups, 2 subjects are used as training dataset and the remaining subjects are used as testing dataset. We generated multi-hop features of 3HGs for each subject and each 3HG was used as a data sample in the training process.

**Model Setting.** For multi-hop features, we generated 1-hop ($l$=1) and 2-hop features ($l$=2) and compared the performance of the model using different multi-hop features. In our experiments, the embedding vectors are initialized by identity matrix to ensure the initial distance between any two embedding vectors are the same. We adopted the embedding dimension $d$ =128. The fusion operation – $W^{Fusion}$, and the two decoder operations – $W^{D_1}$ and $W^{D_2}$, are implemented by fully connected layers and the parameters were initialized following the Xavier scheme. The entire model was trained in an end-to-end manner. The Adam optimizer was used to train the whole model with standard learning rate 0.001, weight decay 0.01, and momentum rates (0.9, 0.999).

### 3.2 Reconstruction Performance

The whole framework is trained by a self-supervised reconstruction task. We compared the reconstruction performance of nine training strategies by mean square error (MSE). For each training strategy, we repeated the experiments for 10 times and summarized the averaged MSE in Fig. 2. Compared to few-shot strategies, the MSE of learning from scratch is dozens of times greater, especially for 2-hop embedding, whose features are more complex than 1-hop features. We also compared different few-shot strategies. Taking 2-hop embedding learning of 34W group as an example (*the zoomed box in (b)*), we used three different few-shot strategies: before transferring to 34W, the models have been trained from scratch by 1) adult group only, 2) adult group and then transfer to 2Y group, and 3) adult group and then successively transfer to 2Y group and 6M group. It is obvious that the models which have been trained by more groups can achieve better performance. It is because using more age groups can provide more common patterns that can be effectively captured by our model. Moreover, our proposed few-shot strategies can converge significantly faster (<30 epochs) than training from scratch only (>300 epochs).

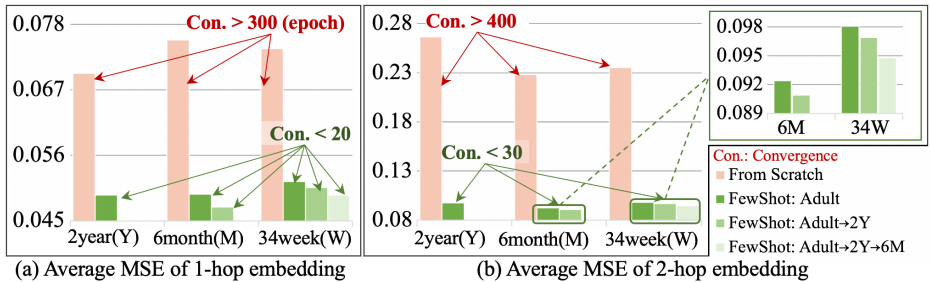

**Fig. 2.** Averaged mean square error (MSE) of reconstruction task. For each few-shot strategy, we repeated the experiments for 10 times and calculated the averaged MSE. (a) Averaged MSE of 1-hop feature. (b) Averaged MSE of 2-hop features.

### 3.3 Embedding Comparison

The learned embeddings directly encode the ROI labels based on the 3HG multi-hop features of the whole population. Effective embeddings should have the capability to represent consistent patterns of 3HG connections. We used the number of multi-hop 3HG connections between two ROIs as ground truth to measure the connection strength within each ROI pair. A larger value means the two ROIs have a stronger connection and their embedding vectors should have high similarity in the embedding space. We adopted cosine similarity to calculate the embedding similarity between each ROI pair and obtained the similarity matrix. To quantitively measure the similarity between embedding matrices and the ground truth matrix, we calculated *structure similarity index measure (SSIM)* between them and showed the results in Fig. 3. For both 1-hop and 2-hop cases, the embeddings learned by few-shot methods have much higher similarity with the ground truth than training from scratch. As the features become more complicated (*from 1-hop features to 2-hop features*), the similarity between embeddings with ground truth dropped dramatically from over 0.5 to 0.07 in the groups of 34week and 6month when training from scratch, while the performance of few-shot embedding is very stable. This result suggests that the proposed few-shot embedding method can effectively recover the anatomical common patterns as well as transfer the learned patterns to other groups.

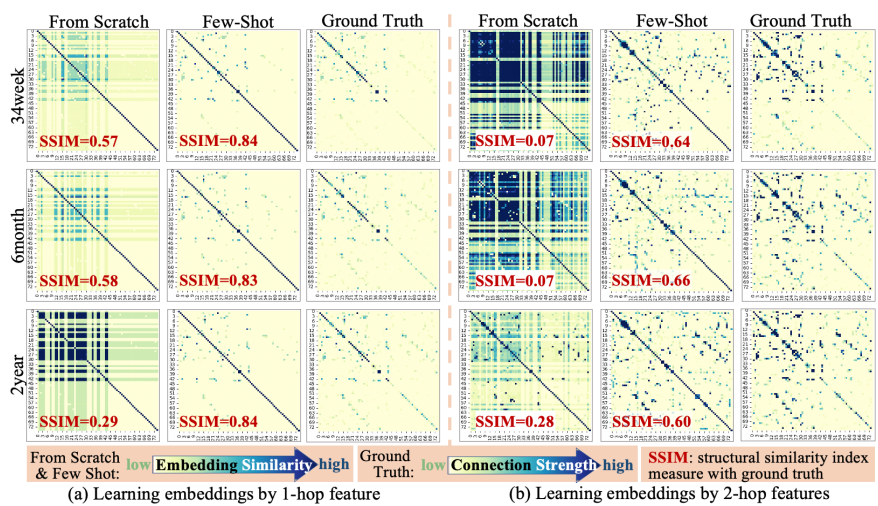

(a) Learning embeddings by 1-hop feature    (b) Learning embeddings by 2-hop features

**Fig. 3.** Evaluation of the learned embeddings. Ground Truth: the number of 3HG connections between ROI pairs. From Scratch/Few-Shot: the similarity between the learned embedding vectors using different learning strategies. For each matrix, the order of the brain regions is the same as the order defined in Destrieux atlas [14].

### 3.4    Embedding of Regularity and Individual Variability

After the model was well-trained, we generated the 3HG embedding according to formular (2). To examine if the learned embeddings can effectively represent regularity/individual variability of anatomical structures, we established the correspondence of 3HGs on different brains based on the embedding similarity. We randomly selected four 3HGs in different brain regions (*pointed by yellow arrows*) to find their corresponding 3HGs in different brains. Here, we consider two 3HGs have correspondence if they have high cosine similarity (>0.9). The results were shown in Fig. 4. The 3HGs with correspondence were organized into the same block. Each one was represented by a bubble that is color coded using the cosine similarity. From the results we can see that our embedding derived correspondences are anatomically meaningful: the joint of specific gyri. Note that due to the widely existing individual variability of folding patterns, one-to-one correspondence may not exist for some brains. In such situation, our proposed embedding method can provide a reliable inference of the complicated many-to-many correspondence (multiple bubbles in the same brain in Fig. 4) by disentangling the regularity from individual variability of brain anatomical structures.

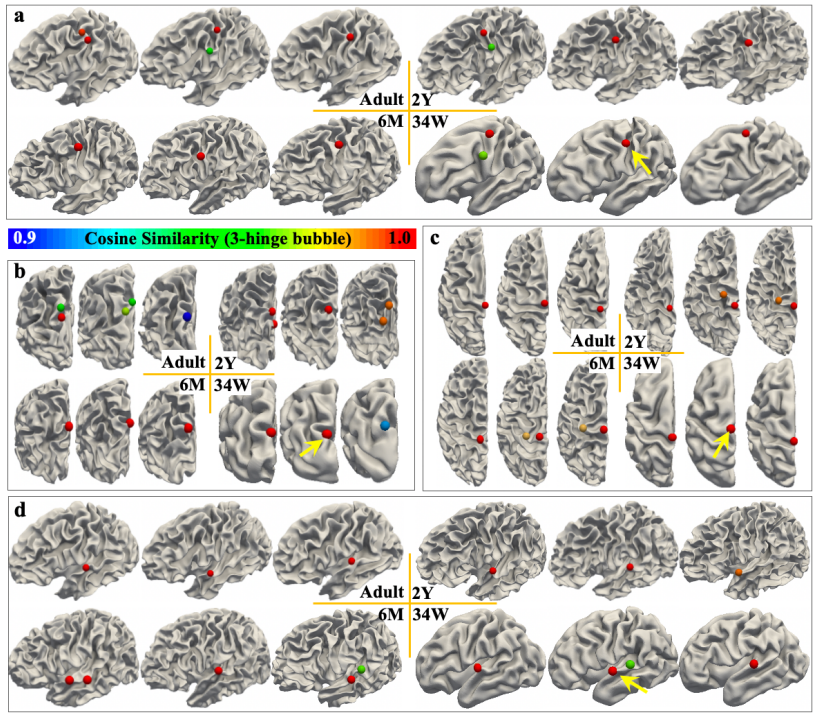

**Fig. 4.** Results of 3HG correspondence. The correspondence of 3HGs can be calculated by embedding similarity. We randomly selected four 3HGs (*pointed by yellow arrows*) to find their corresponding 3HGs (*cosine similarity over 0.9*) in different brains. Four groups of corresponding 3HGs of 3 subjects from 4 age groups are showed in four blocks. Each 3HG was represented by a bubble and color coded by the cosine similarity.

## 4 Conclusion

In this work, we proposed a few-shot embedding framework to represent the regularity and variability of folding patterns for different age groups. We described the brain anatomy by 3HG network and learned a group of feature embeddings to capture the regularity of the folding pattern. The variability of landscapes is embedded by a group of specific multi-hop feature coefficients. Experimental results suggest that the proposed embedding method can provide a reliable inference of the complicated many-to-many correspondence and represent the regularity and variability of folding pattern. Moreover, by introducing the few-shot learning, the proposed framework can generalize well in the age group of early neurodevelopment with very limited samples.